\documentclass[pteplogo,hdvipdfm]{ptephy_v1}

\preprintnumber{XXXX-XXXX} 

\usepackage{amssymb}
\usepackage{amsmath}
\usepackage[all, warning]{onlyamsmath}

\newcommand{\pdel}[2]{\frac{\partial #1}{\partial #2}}
\newcommand{\ptdel}[2]{{\partial #1}/{\partial #2}}
\newcommand{\vev}[1]{\langle #1 \rangle}
\newcommand{\diracslash}[1]{#1 \hspace{-5pt}/ \:}
\newcommand{\Pslash}{P \hspace{-7pt}/ \:}

\usepackage[normalem]{ulem} 
\usepackage{color}

\begin{document}

\title{Role of U$_{A}$(1) breaking term in dynamical chiral symmetry breaking of chiral effective theories}


%
%

\author[1,2]{Shinnosuke~Kono}
\affil{Department of Physics, Tokyo Metropolitan University, Tokyo 192-0397, Japan}

\author[2,1]{Daisuke~Jido}
\affil{Department of Physics, Tokyo Institute of Technology, Tokyo 152-8551, Japan  
}

\author[3]{Yoshiki~Kuroda}
\author[3]{Masayasu~Harada} 
\affil{Department of Physics, Nagoya University, Nagoya, 464-8602, Japan}



\begin{abstract}
The spontaneous breaking of chiral symmetry is examined by 
chiral effective theories, such as the linear $\sigma$ model 
and the Nambu--Jona-Lasinio (NJL) model. 
We discuss the properties of the sigma meson regarded as the quantum fluctuation 
of the chiral condensate when the chiral symmetry is spontaneously broken mainly by
the U$_{A}$(1) anomaly. 
We derive a mass relation among the SU(3) flavor singlet mesons, $\eta_{0}$ and $\sigma_{0}$,
in the linear $\sigma$ model to be satisfied 
for the anomaly driven symmetry breaking in the chiral limit, 
and find that it is also supported in the NJL model. 
With the explicit breaking of chiral symmetry, we 
show that both 
the chiral effective models reproducing the observed physical quantities suggest that the $\sigma_{0}$ meson 
should have a mass smaller than an order of 800 MeV when the anomaly 
driven symmetry breaking takes place. 
\end{abstract}

\subjectindex{
D32,
B60 
}

\maketitle

\section{Introduction}

Chiral symmetry and its dynamical breaking are the most important
ingredients of quantum chromodynamics (QCD).  
There are several effective theories which model chiral symmetry in QCD.
In the linear $\sigma$ 
model~\cite{GellMann:1960np}, chiral symmetry is realized by the scalar and pseudoscalar 
mesons, which are transformed each other by the chiral transformations. 
The emergence of the vacuum expectation values of  
the scalar fields with isospin $I=0$ breaks
the chiral symmetry spontaneously,
and the pseudoscalar mesons become the Nambu-Goldstone bosons associated 
with the spontaneous breaking.
In standard $\sigma$ models, a negative coefficient of the quadratic term
of the self-energy of the scalar field induces the spontaneous chiral symmetry 
breaking. In the Nambu--Jona-Lasinio (NJL) model~\cite{Nambu:1961tp}, 
chiral symmetry is defined for the quark fields in the same way as QCD, 
and a sufficiently strong four-point coupling of the quark fields breaks 
chiral symmetry dynamically and the quark masses are generated. 

Quantum anomaly~\cite{Weinberg:1975ui,Witten:1979vv,Veneziano:1979ec} 
is also one of the most important phenomena in QCD. 
The quark triangle loop breaks the U(1)$_{A}$ symmetry 
for the quark field explicitly~\cite{Adler:1969gk,Bell:1969ts,Bardeen:1969md}, 
and it makes the $\eta^{\prime}$ meson fail to become a Nambu-Goldstone boson
associated with the dynamical breaking of the three-flavor chiral symmetry~\cite{tHooft:1986nc}.
Such U(1)$_{A}$ symmetry breaking can be also implemented to 
the chiral effective theories, for example, by the determinant type
Kobayshi-Maskawa-'t~Hooft
interaction~\cite{Kobayashi:1970ji,Kobayashi:1971qz,'tHooft1976,tHooft:1976rip,Rosenzweig:1979ay,DiVecchia:1980yfw,Witten:1980sp}. 
The dynamical breaking of chiral symmetry and the U(1)$_{A}$ anomaly have been discussed 
also by the instanton liquid model~\cite{Schafer:1996wv}.

The properties of the light scalar meson, $\sigma$, are also interesting issues. 
There must exist the scalar fields which transform into the pion fields under the chiral transformation, {\it i.e.}\ the chiral partners of the pions.
It is not well known yet how such scalar fields appear 
in the physically observed spectra.
There are a lot of
interpretations of their structure,
such as two-quark $\bar qq$, tetraquark $\bar q \bar q qq$, glueball, 
two-pion or pion-kaon correlation and their mixture
(see e.g., Refs.~\cite{Black:1998wt,Ishida:1999qk,Naito:2002,Kunihiro:2003yj,Pennington:2007yt,Klempt:2007cp,Fariborz:2009cq,Hyodo:2010jp,Parganlija:2012fy,Pelaez:2015qba,Pelaez:2016klv,Achasov:2017ozk}). 
Certainly it is very important to know 
the properties of 
the quantum fluctuation of the chiral condensates.

In this article, 
by using the linear $\sigma$ model~\cite{Schechter1971,Kawarabayashi1980,Ishida:1999qk,Lenaghan2000,Schaefer:2008hk} and
the NJL model~\cite{Bernard:1987gw,Kunihiro:1987bb,Vogl:1991qt,Hatsuda:1994pi,Takizawa:1996nw} for three flavors with the anomaly contributions, 
we see that, despite the situation that chiral symmetry is not broken 
spontaneously without the anomaly term, 
chiral symmetry is to be broken spontaneously, if the anomaly contribution is large enough.
We examine the condition on the mass of the flavor singlet scalar meson $\sigma_{0}$ and compare
the $\eta^{\prime}$ mass 
with the chiral symmetry breaking mass scale when
such an anomaly driven symmetry breaking takes place. 
The spontaneous chiral symmetry breaking plays a significant role also for
the generation of the $\eta^{\prime}$ mass~\cite{Lee:1996zy,Jido:2011pq}.

In this paper we assume that $\eta^{\prime}$ is a flavor singlet 
and do not introduce the mixing with the octet $\eta_{8}$ for simplicity,
encouraged by the fact that the mixing angle between $\eta_{0}$ and $\eta_{8}$
is found to be small thanks to the large mass difference of the singlet and octet
isoscalar pseudoscalar mesons.  
In order to discuss the quantitative feature of the mass scale 
of the scalar mesons, 
we do not consider the flavor mixing in the scalar fields, either. 
For more quantitative analyses one should introduce the mixing 
in both scalar and pseudoscalar mesons. 

This paper is organized as follows. In Sec.\ 2, using the linear $\sigma$ model 
we show that the anomaly driven symmetry breaking takes place with a sufficiently 
large anomaly contribution and find 
the condition for the $\sigma_{0}$ mass 
in the anomaly driven breaking. In Sec.~3, 
we confirm that the NJL model provides
the consistent consequences with the linear $\sigma$ model.
Section 4 is devoted to summary of this article. 

\section{Linear $\sigma$ model}
Let us start with introducing the linear $\sigma$ model for three flavors with the U$_{A}$(1) breaking term. 
The Lagrangian of the linear $\sigma$ model that we consider here is 
given in Refs.~\cite{Sakai:2013nba,Sakai:2016vcl} as
\begin{align}
   {\cal L}_{\sigma} = &
   \frac 12 \textrm{ Tr} [\partial_{\mu} \Phi \partial^{\mu} \Phi^{\dagger}] 
   - \frac{\mu^{2}}{2} \textrm{ Tr} [ \Phi\Phi^{\dagger}]    
   - \frac{\lambda}{4} \textrm{ Tr} [ (\Phi\Phi^{\dagger})^{2}] 
   - \frac{\lambda^{\prime}}{4} \left(\textrm{ Tr} [\Phi\Phi^{\dagger}] \right)^{2}
   \nonumber \\ &
   + A \textrm{ Tr}[ \chi \Phi^{\dagger} + \chi^{\dagger} \Phi] 
   + \sqrt 3 B (\det \Phi + \det \Phi^{\dagger}). \label{Lsig}
\end{align}
We consider the meson fields $\Phi$ and $\Phi^{\dagger}$ to belong 
to the $( {\bf 3},\bar {\bf 3}) \oplus (\bar{\bf 3},  {\bf 3})$
representation of the SU(3)$_{L}\otimes$SU(3)$_{R}$ chiral group,
under which the meson fields $\Phi$ and $\Phi^{\dagger}$ transforms as
\begin{equation}
   \Phi \to L \Phi R^{\dagger}, \qquad
   \Phi^{\dagger} \to R \Phi^{\dagger} L^{\dagger}
\end{equation}
with the SU(3)$_{L}$ and SU(3)$_{R}$ transformations, $L$ and $R$, respectively. 
The external scalar field $\chi$ is supposed to transform as $\chi \to L \chi R^{\dagger}$
under the SU(3)$_{L}$ and SU(3)$_{R}$ symmetry, and thus
the Lagrangian~\eqref{Lsig} is invariant under these transformations.

In the $(\bar {\bf 3}, {\bf 3}) \oplus ({\bf 3}, \bar {\bf 3})$ multiplet, 
the scalar and pseudoscalar mesons in the singlet and octet representations of SU(3)$_{V}$, 
18 mesons in total, are included as shown in Ref.~\cite{Jido:2011pq}. 
The meson field $\Phi$ is written with the scalar meson field $\Phi_{s}$ 
and the pseudoscalar meson field $\Phi_{p}$ as 
\begin{equation}
 \Phi=\Phi_s+i\Phi_{p}=\sum^8_{a=0}\frac{\lambda_a\sigma_a}{\sqrt{2}}+i\sum_{a=0}^8\frac{\lambda_a\pi_a}{\sqrt{2}},
\end{equation}
where $\lambda_{a}$ $(a=1,\dots, 8)$ are the Gell-Mann matrices and $\lambda_{0} = \sqrt{\frac 23} {\bf 1}$
with the unit matrix~$\bf 1$. These matrices are normalized as $\textrm{ Tr}(\lambda_{a} \lambda_{b}) = 2 \delta_{ab}$
for $a,b = 0,1, \dots, 8$. 
The scalar and pseudoscalar fields, $\sigma_{a}$ and $\pi_{a}$, are chiral partners 
and transform each other under the chiral rotations. 
The external scalar field $\chi$ is given by scalar and 
pseudoscalar fields, $S$ and $P$, as $\chi = S + i P$ and 
they are fixed as $S = {\cal M} \equiv \textrm{ dia}(m_{q}, m_{q}, m_{s})$ and $P=0$.
The finite quark masses, $m_{q} \neq 0$, $m_{s} \neq 0$, introduces 
the explicit chiral symmetry breaking.
We assume the isospin symmetry by $m_{q} = m_{u} = m_{d}$, and 
with a heavier strange quark mass $m_{s}  > m_{q}$ the flavor symmetry 
is broken.
The term with coefficient $B$
breaks the U$_{A}$(1) symmetry explicitly. 

In the mean field approximation at the tree level, the effective potential is 
given by 
\begin{align}
   V = &   \frac{\mu^{2}}{2} \textrm{ Tr} [ \Phi\Phi^{\dagger}]     
   + \frac{\lambda}{4} \textrm{ Tr} [ (\Phi\Phi^{\dagger})^{2}]  
   + \frac{\lambda^{\prime}}{4} \left(\textrm{ Tr} [\Phi\Phi^{\dagger}] \right)^{2}
   - A \textrm{ Tr}[ {\cal M} ( \Phi^{\dagger} +  \Phi )] 
   \nonumber \\ &
   - \sqrt 3 B (\det \Phi + \det \Phi^{\dagger}).
\end{align}
For the stability of the system, we assume $\lambda >0$ and $\lambda^{\prime} >0$. 
The vacuum is defined by the minimum of the effective potential. 

Let us first consider the chiral limit with $m_{q} = m_{s} = 0$. 
The system has no explicit breaking of chiral symmetry and 
possesses the SU(3) flavor symmetry. 
The effective potential in the chiral limit is calculated as
\begin{equation}
  V = \frac12 \mu^{2} \vev{\sigma_{0}}^{2} + \frac14(\frac{\lambda}{3} + \lambda^{\prime}) \vev{\sigma_{0}}^{4} 
  - \frac23 B \vev{\sigma_{0}}^{3}.
\end{equation}
The vacuum condition is obtained by
$\pdel{V}{\sigma_{0}} = 0$ at $\sigma_{0} = \langle \sigma_{0} \rangle$ 
and the other meson fields vanishing as
\begin{equation}
  \langle \sigma_{0} \rangle  \left[  \mu^{2}  
   + ( \frac{\lambda}{3}+\lambda^{\prime}) \langle \sigma_{0} \rangle^{2}
   - 2 B \langle \sigma_{0} \rangle \right] =0 .
\end{equation}
This equation has a trivial solution $\vev{\sigma_{0}} = 0$. 
Nontrivial solutions, $\vev{\sigma_{0}} \neq 0$, are also possible 
if the discriminant of the quadratic equation 
\begin{equation}
\Delta = B^{2} - \mu^{2}( \lambda/3 + \lambda^{\prime})
\end{equation}
is positive.  
In the case of $B=0$, for 
$\mu^{2}>0$, only the trivial solution is possible
because of $\Delta < 0$ due to  $\lambda >0$ and $\lambda^{\prime} >0$. 
Thus, chiral symmetry is not broken. 
For $\mu^{2} < 0$, because the discriminant is always positive, $\Delta >0$, 
and the system with $\vev{\sigma_{0}} =0$ is unstable,
the vacuum is realized with a finite $\vev{\sigma_{0}}$,
which breaks chiral symmetry spontaneously. 
In this way, the sign of $\mu^{2}$ controls the nature of the vacuum. 
This is a conventional understanding for the spontaneous breaking of chiral symmetry
in the linear $\sigma$ model. 

Here let us consider the situation that chiral symmetry is spontaneously broken
with $\mu^{2}>0$ and $B \neq 0$. Even though $\mu^{2} > 0$,
with appropriate values of $B$ satisfying $B^{2} > \mu^{2}(\lambda/3+\lambda^{\prime})$, 
the discriminant $\Delta$ can be positive and 
one finds solutions of the vacuum condition:
\begin{equation}
   \vev{\sigma_{0}} = \frac{B \pm \sqrt {\Delta}}{\lambda/3 + \lambda^{\prime}}.
   \label{vcon0}
\end{equation}
There are two solutions for this quadratic equation.  
The effective potential has a local minimum at  
the solution with the $+$ sign in Eq.~\eqref{vcon0} and 
a local maximum at the other solution.  
For $\mu^{2}>0$, the effective potential has a local minimum also
at $\langle \sigma_{0} \rangle =0$. 
To realize the vacuum with $\langle \sigma_{0} \rangle \neq 0$, 
the effective potential has to be 
the minimum there. 
The value of the effective potential at the vacuum $\sigma_{0} = \langle \sigma_{0} \rangle$ is calculated as
\begin{equation}
   \left. V \right|_{\sigma_{0} = \langle \sigma_{0} \rangle} 
   = \frac{1}{4} \mu^{2} \langle \sigma_{0} \rangle^{2} - \frac16 B \langle \sigma_{0} \rangle^{3},
\end{equation}
where we have used  Eq.~\eqref{vcon0} with $+$ sign. To have the minimum at $\sigma_{0} = \langle \sigma_{0} \rangle$,  
$ \left. V \right|_{\sigma_{0} = \langle \sigma_{0} \rangle} <   \left. V \right|_{\sigma_{0} =0} = 0 $, that is, 
\begin{equation}
   ( \ 0 <\  )\quad
   \mu^{2} < \frac 23 B \langle \sigma_{0} \rangle. \label{ADScon}
\end{equation}
In this way, as pointed out in Ref.~\cite{Lenaghan2000}, $\mu^{2}<0$ is not a condition
of the spontaneous breaking any more if the effect of the anomaly term is sufficiently large.  
We call Eq.~\eqref{ADScon} anomaly driven solution.
Even though the parameters $\mu^{2}$ and $B$ are not controllable parameters in 
nature governed by QCD, here we just consider such situations to discuss its implications. 

We express the condition for the anomaly driven solution with physical observables. 
The masses of $\sigma_{0}$ and $\eta_{0}$ can be calculated at the tree level 
as a curvature of the effective potential at the vacuum point: 
\begin{align}
   m_{\sigma_{0}}^{2} & 
   = \mu^{2} + (\lambda + 3 \lambda^{\prime} ) \langle \sigma_{0} \rangle^{2} - 4B \langle \sigma_{0} \rangle 
   =- 2\mu^{2} + 2 B \langle \sigma_{0} \rangle,
   \label{sig0mass0} \\
  m_{\eta_{0}}^{2}  & 
   = \mu^{2} + (\frac{\lambda}{3} +\lambda^{\prime} ) \langle \sigma_{0} \rangle^{2}   + 4B \langle \sigma_{0} \rangle
   = 6 B \langle \sigma_{0} \rangle,
    \label{eta0mass0}
\end{align}
with Eq.~\eqref{vcon0}. 
From Eqs.~\eqref{sig0mass0} and \eqref{eta0mass0}, we find a relation 
\begin{equation}
   6 \mu^{2} = m_{\eta_{0}}^{2} - 3 m_{\sigma_{0}}^{2}.  \label{murel}
\end{equation}
Therefore, for the anomaly driven solution which satisfies 
Eq.~\eqref{ADScon},
one should have
\begin{equation}
\tfrac19 m_{\eta_{0}}^{2} < m_{\sigma_{0}}^{2} <  \tfrac13 m_{\eta_{0}}^{2}, \label{eq:ineq}
\end{equation}
which implies that the $\sigma_{0}$ mass should be small.
Reference \cite{Schaefer:2008hk} also found that for smaller $\sigma$ masses with the 
anomaly term the parameter $\mu^{2}$ becomes positive, but there it was incorrectly concluded that in such a case spontaneous symmetry breaking would be lost and the condensates would vanish in the chiral limit. 
The inequality \eqref{eq:ineq}
can be understood as follows. The $\eta_{0}$ mass is given by the anomaly parameter $B$
in the chiral limit, while the $\sigma_{0}$ mass is determined also by $\mu^{2}$. In the anomaly driven 
solution, the anomaly term plays a more significant role than the $\mu^{2}$ term.  For that reason,
the parameter $B$ has a larger value and the $\eta_{0}$ mass should be sufficiently large. 

It is also interesting that 
the usual spontaneous breaking takes place for the case of 
\begin{equation}
m_{\eta_{0}}^{2} < 3 m_{\sigma_{0}}^{2}.
\end{equation} 
This implies that the $\sigma_{0}$ mass should be sufficiently large for the standard spontaneous breaking.
Here the $\sigma_{0}$ field is 
a quantum fluctuation mode of the condensate which breaks the chiral symmetry spontaneously, {\it i.e.}\ a chiral partner of the Nambu-Goldstone boson. 
The actual scalar meson can be a mixture of $\sigma_{0}$ and other components.

\begin{figure}[t]
\centering
   \includegraphics[width=0.8\linewidth]{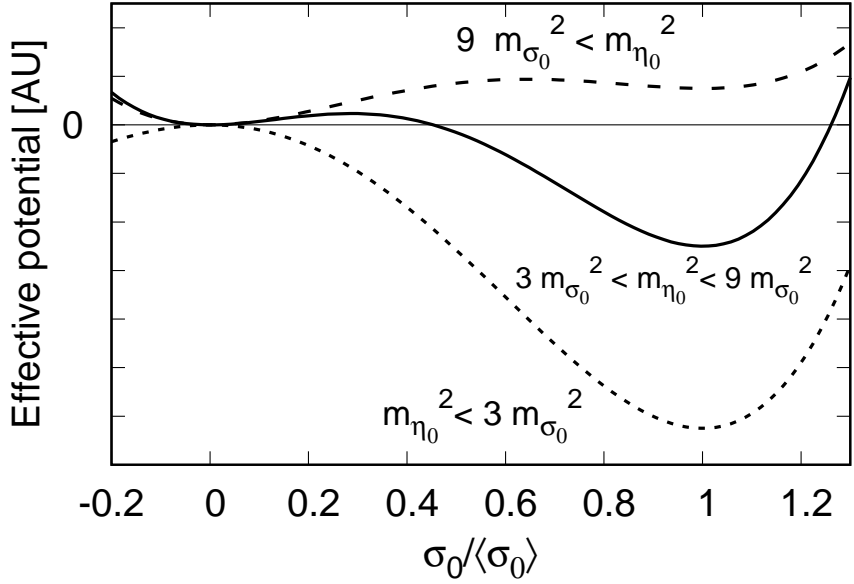}
\caption{Schematic plot of the effective potential of the linear $\sigma$ model 
in the chiral limit as a function of the $\sigma_{0}$ mean field scaled 
to the vacuum expectation value $\vev{\sigma_{0}}$,
$\sigma_{0}/\vev{\sigma_{0}}$, in arbitrary units. The dotted, solid and dashed lines
are calculations with $m_{\eta_{0}}^{2}< 3 m_{\sigma_{0}}^{2} $,  
$3 m_{\sigma_{0}}^{2} <  m_{\eta_{0}}^{2} < 9 m_{\sigma_{0}}^{2}$ and 
$9 m_{\sigma_{0}}^{2} < m_{\eta_{0}}^{2}$, respectively,
(these masses are evaluated at the vacuum with the finite $\vev{\sigma_{0}}$). 
The standard breaking is for 
the case of the dotted line, while the anomaly driven breaking takes place with the case of 
solid line.} \label{fig:sigpot}
\end{figure}

In Fig.~\ref{fig:sigpot}, we show a schematic plot of the effective potential $V$
for three typical cases. 
The dotted line presents the standard spontaneous breaking. At the origin 
the effective potential has a local maximum and the chiral symmetry 
is spontaneously broken with $\vev{\sigma_{0}} \neq 0$. 
The solid line shows the effective potential for the 
anomaly driven solution, where thanks to $\mu^{2}>0$ the effective potential 
is convex downward at $\sigma_{0}=0$ but it has the minimum at  $\sigma_{0}\neq 0$.
As shown in the figure, for the case of the anomaly driven solution, the 
effective potential has a local minimum at the origin and this situation could indicate 
strong first order
phase transition at finite temperature and/or density. 
If the $\sigma_{0}$ mass 
evaluated at the vacuum with the finite $\vev{\sigma_{0}}$
is sufficiently small, {\it i.e.} $m_{\sigma_0}^2 < m_{\eta_0}^2/9$, 
the chiral symmetry is not spontaneously broken.

For more quantitative discussion, we introduce the finite quark masses;
$m_{s} > m_{q} \neq 0$.  The flavor SU(3) breaking allows us to have 
a finite value of $\langle \sigma_{8} \rangle$. 
The vacuum conditions are obtained by 
$
     \ptdel{V}{\sigma_{0}} = \ptdel{V}{\sigma_{8}} = 0
$
with $\sigma_{0} = \vev{\sigma_{0}}$, $\sigma_{8} = \vev{\sigma_{8}}$ and 
vanishing other meson fields. 

Here we would like to derive the extension of the relation~\eqref{murel} off the chiral limit. 
For a finite $\vev{\sigma_{0}}$, the vacuum conditions
read
\begin{align}
   \mu^{2} + \frac \lambda 3& \vev{\sigma_{0}}^{2}(1 + 6 \epsilon^{2} - 2 \epsilon^{3})
   + \lambda^{\prime} \vev{\sigma_{0}}^{2}(1 + 2 \epsilon^{2})  
    - 2 A\frac{2m_{q}+m_{s}}{\vev{\sigma_{0}}} - 2B \vev{\sigma_{0}} (1 - \epsilon^{2}) = 0,  \label{vcon0oc}\\
 \mu^{2}\epsilon+ \lambda & \vev{\sigma_{0}}^{2}(\epsilon - \epsilon^{2} + \epsilon^{3})
   + \lambda^{\prime} \vev{\sigma_{0}}^{2}(\epsilon + 2 \epsilon^{2})
  - 2 A\frac{m_{q}-m_{s}}{\vev{\sigma_{0}}} + 2B \vev{\sigma_{0}} (\epsilon + \epsilon^{2}) = 0, \label{vcon8oc}
\end{align}
from $\ptdel{V}{\sigma_{0}} = 0$ and $ \ptdel{ V}{ \sigma_{8}} = 0$, respectively. 
Here we have introduced the flavor SU(3) breaking parameter $\epsilon$ as
\begin{equation}
   \epsilon = \frac{\vev{\sigma_{8}}}{\sqrt 2 \vev{\sigma_{0}}}.
\end{equation}
At the tree level, the condensates are written in terms of the meson decay constants as
\begin{align}
   \vev{\sigma_{0}} &= \frac{1}{\sqrt 6} (f_{\pi} + 2 f_{K}), \\
   \vev{\sigma_{8}} & = \frac 2 {\sqrt 3}     (f_{\pi} - f_{K}),
\end{align}
where $f_{\pi}$ and $f_{K}$ are the pion and kaon decay constants, respectively,
and the meson decay constants are defined by $\langle 0 | A_{\mu}^{a}(x) | \pi^{b}(p) \rangle = i \delta_{ab} p_{\mu}f_{b} e^{-ip\cdot x}$. The axial current $A_{\mu}^{a}(x)$ is obtained as the Noether current associated with 
the axial transformation and calculated as $A_{\mu}^{a} = \textrm{Tr}[\partial_{\mu} \Phi_{p} \{ \lambda^{a}, \Phi_{s}\}
- \partial_{\mu} \Phi_{s} \{ \lambda^{a}, \Phi_{p}\}]$. 
Using observed values $f_{\pi} = 92.2$~MeV and $f_{K} = 110.4$~MeV, we determine 
the values of the condensates as $\vev{\sigma_{0}} = 128$~MeV 
and $\vev{\sigma_{8}} = -21.0$~MeV~\cite{Sakai:2013nba}. Thus, we find the value of $\epsilon$ as
\begin{equation}
   \epsilon = -0.116. \label{eps}
\end{equation}

The masses are calculated as
\begin{subequations}
\label{mesonmasses}
\begin{align}
  m_{\sigma_{0}}^{2}  = \ & \mu^{2} + \lambda \vev{\sigma_{0}}^{2} ( 1 + 2 \epsilon^{2}) 
  + \lambda^{\prime} \vev{\sigma_{0}}^{2} (3 + 2\epsilon^{2}) - 4 B \vev{\sigma_{0}},  \label{sig0mass} \\
  m_{\eta_{0}}^{2} = \ &  \mu^{2} + \frac \lambda 3 \vev{\sigma_{0}}^{2}(1 + 2 \epsilon^{2}) 
  + \lambda^{\prime} \vev{\sigma_{0}} (1 + 2 \epsilon^{2}) + 4 B \vev{\sigma_{0}}, \label{eta0mass} \\
  m_{\pi}^{2} = \ & \mu^{2} + \frac \lambda 3 \vev{\sigma_{0}}^{2} ( 1 + 2 \epsilon + \epsilon^{2}) 
  + \lambda^{\prime} \vev{\sigma_{0}}^{2} (1 + 2 \epsilon^{2}) - 2B \vev{\sigma_{0}} (1 - 2 \epsilon) , \label{pimass}\\
  m_{\eta_{8}}^{2} = \ & \mu^{2} + \frac \lambda 3 \vev{\sigma_{0}}^{2} (1 - 2 \epsilon + 3 \epsilon^{2}) 
  + \lambda^{\prime} \vev{\sigma_{0}}^{2} (1 + 2 \epsilon^{2}) -2 B \vev{\sigma_{0}} (1 + 2 \epsilon). \label{eta8mass}
\end{align}
\end{subequations}

From Eqs.~\eqref{vcon0oc}, \eqref{vcon8oc}
and \eqref{mesonmasses}, 
we obtain the relation 
\begin{align}
6\mu^{2}\left(1-\tfrac{1}{2}\epsilon\right) =\  &
m_{\eta_{0}}^{2}\left(1-\tfrac{1}{2}\epsilon-6\epsilon^{2}+\epsilon^{3}-8\epsilon^{4}\right) 
-3m_{\sigma_{0}}^{2}\left(1-\tfrac{1}{2}\epsilon+2\epsilon^{2}-\epsilon^{3}\right)  \nonumber \\
& +m_{\eta_{8}}^{2}\left(4-5\epsilon+6\epsilon^{2}-8\epsilon^{3}+4\epsilon^{4}\right) 
  +m_{\pi}^{2}\left(4+\epsilon+6\epsilon^{2}+4\epsilon^{3}+4\epsilon^{4}\right),
\end{align}
and with Eq.~\eqref{eps} 
\begin{equation}
   6\mu^{2}=0.92m_{\eta_{0}}^{2}-3.08m_{\sigma_{0}}^{2}+4.42m_{\eta_{8}}^{2}+3.96m_{\pi}^{2}.
\end{equation}
Using empirical values of the meson masses,
$m_{\eta^{\prime}}=958$~MeV, $m_{\eta}=550$~MeV, $m_{\pi}=140$~MeV, 
where we omit to resolve the mixing of $\eta_{0}$ and $\eta_{8}$
thanks to the small mixing angle, we find that 
\begin{align}
   m_{\sigma_{0}}& < 840\ \textrm{ MeV}, \qquad  \textrm{ for}\quad \mu^{2} > 0, 
   \label{eq:LsMano} \\
   m_{\sigma_{0}}& > 840\ \textrm{ MeV}, \qquad  \textrm{ for}\quad \mu^{2} < 0 .   
   \label{eq:LsMstd}
\end{align}
A similar result has been already obtained in Ref.\cite{Schaefer:2008hk}.
Equation \eqref{eq:LsMano} implies that, if the anomaly driven breaking of chiral symmetry 
is the case in nature, 
the mass of the $\sigma_{0}$ meson, 
which is the quantum fluctuation mode of the chiral condensate,
should be smaller than 840~MeV. This is very interesting, because, 
if the light scalar resonance appearing around 500~MeV in the $\pi\pi$ 
scattering with $I=0$ is the chiral partner of the pion, 
that is the above $\sigma_0$ meson,
or at least 
it has a large contribution of the 
$\sigma_0$ meson, the anomaly driven 
chiral symmetry breaking should be the case.
Alternatively if we could 
rule out the anomaly driven solution from some reason, we would 
have the lower limit of the mass of the $\sigma_0$ meson. 

\section{Nambu Jona-Lasinio model}
Let us consider the Nambu Jona-Lasinio model (NJL model) 
for the three flavors with the anomaly term~\cite{Kunihiro:1987bb}. The Lagrangian is 
given as
\begin{align}
  {\cal L}_\textrm{ NJL} =& \
   \bar \psi ( i \diracslash{\partial} -{\cal M}) \psi
    + g_{S} \sum_{a=0}^{8}\left[ \left(\bar \psi \frac{\lambda_{a}}2 \psi\right)^{2} 
    + \left(\bar \psi i \gamma_{5} \frac{\lambda_{a}}2 \psi\right)^{2} \right] 
    \nonumber \\ & \ 
    + \frac{g_{D}}2 \left\{ \det[ \bar \psi_{i} (1+\gamma_{5}) \psi_{j}]
    + \det[ \bar \psi_{i} (1-\gamma_{5}) \psi_{j}] \right\}
\end{align}
for the quark field $\psi = (u,d,s)^{T}$, where $\cal M$ is the quark mass matrix
given by ${\cal M} = \textrm{ diag}(m_{q}, m_{q}, m_{s})$ with assuming
isospin symmetry $m_{q} = m_{u} = m_{d}$, $\lambda^{a}$
($a=0,1, \dots, 8$) are
the Gell-Mann matrices,
$\det$ is determinant 
taken in the flavor space, 
and $g_{S}$ and $g_{D}$ are the coupling constants for the four-point vertex and 
the determinant-type U$_{A}$(1) breaking term, respectively. 
These coupling constants are dimensionful. 

In the mean field approximation, the dynamical quark masses, 
$M_{q}$ and $M_{s}$,
are calculated by the gap equations:
\begin{subequations} \label{gapeq}
\begin{align}
   M_{q} &= m_{q} - g_{S} \vev{\bar qq} - g_{D} \vev{\bar qq}\vev{\bar ss}, \label{gapq} \\
   M_{s} &= m_{s} - g_{S} \vev{\bar ss} - g_{D} \vev{\bar qq}^{2},  \label{gaps}
\end{align}
\end{subequations}
where $\vev {\bar qq} = \vev{\bar uu} = \vev{\bar dd}$ with the isospin symmetry. 
The quark condensate is evaluated by the quark propagator 
$S_{F}(x) = -i \vev {0| T [ q(x) \bar q(x)] | 0}$ with the dynamical mass $M_{q}$
as
\begin{equation}
   \vev{\bar qq} = - i N_{c} {\lim_{x \to 0}} \textrm{Tr} [S_{F}(x)], \label{qq}
\end{equation}
for the number of color $N_{c} =3$ and $\textrm{Tr}$ taken for the Dirac indices. 
In our formulation we use the three momentum cutoff $\Lambda$ to calculate 
the momentum integral in $ {\lim_{x \to 0}} \textrm{Tr} [S_{F}(x)]$ by following 
the suggestion given in Ref.~\cite{Hatsuda:1994pi}. 
The three-momentum cutoff $\Lambda$ restricts 
the momentum of virtual quarks not the energy.
The explicit form of Eq.~\eqref{qq} with the cutoff~$\Lambda$
is given in Eq.~\eqref{eq:SF}. 
In this work the cutoff is fixed at $\Lambda = 602.3$ MeV as an example. 
To investigate the role of the anomaly term as a theoretical interest, 
we calculate physical quantities by changing the coupling constants $g_{S}$ and $g_{D}$. 
In order to compare these contributions, we should fix the model by using a certain regularization scheme.

The effective potential in the mean field approximation is given by
\begin{align}
  V(\hat M_{q}, \hat M_{s})  &= N_{c} \sum_{f=q,q,s} i \int \frac{d^{4}p}{(2\pi)^{4}}
   \textrm{Tr} \left[ \log(\diracslash{p}-\hat M_{f})+ \frac{\diracslash{p}-m_{f}}{\diracslash{p} - \hat M_{f}} \right] 
   \nonumber \\ &
   \quad - \frac{g_{S}}{2}(2 \vev{\bar qq}^{2} +  \vev{\bar ss}^{2} )
   - g_{D} \vev{\bar qq}^{2} \vev{\bar ss}. \label{eq:Veff}
\end{align}
This is a function of the dynamical quark masses $\hat M_{q}$ and $\hat M_{s}$. 
The explicit form of the first term of Eq.~\eqref{eq:Veff} is given in Eq.~\eqref{eq:log}. 
The stationary points against the variation of the quark masses are given by 
the gap equations \eqref{gapeq}. 

The mesons are dynamically generated in 
the scattering amplitude $T$ for 
the two-body $\bar q_{i} q_{j}$ channel having the same quantum number as the meson
and obtained as a pole of the scattering amplitude $T$.
The $T$-matrix is calculated
as a solution of the Bethe-Salpeter equation 
\begin{equation}
   T = K +  K J T, \label{BSeq}
\end{equation}
where $J$ is the quark loop function. The explicit forms of the
pseudoscalar and scalar channels are given in Eqs.~\eqref{eq:JPS},
\eqref{eq:JS} and \eqref{eq:JK},
and the loop functions for each channel are given in Eqs.~\eqref{loopfun}.
In this work we use the ladder approximation to calculate 
the Bethe-Salpeter equation~\eqref{BSeq}.
The loop function is calculated using the three momentum cutoff with the same
value of $\Lambda$ as in the calculation of the quark condensate.
The interaction kernels $K$ for the
four-quark interaction are obtained from the determinant-type 
six-point vertex with insertion of one quark condensate 
as well as the four-point vertex
according to the mean field approximation. 
The explicit forms of the interaction kernels 
for the $\eta_{0}$, $\sigma_{0}$, $\pi$, $K$
and $\eta_{8}$ are listed in Eqs.~\eqref{kernel}.
Here we assume the quark contents of $\eta_{0}$ and $\sigma_{0}$ 
to be the flavor singlet $\frac1{\sqrt 3} (\bar uu + \bar dd + \bar ss)$ and 
that of $\eta_{8}$ to be the flavor octet$\frac 1{\sqrt 6}(\bar uu + \bar dd - 2\bar ss)$. 
The pole position of the $T$-matrix can be found by solving
\begin{equation}
   1 - KJ(\sqrt s) = 0, \label{eq:mass}
\end{equation}
for the single channel case. 
Bound states are found below the threshold of the quark and antiquark production
in the first Riemann sheet of the complex $\sqrt s$ plane, 
while virtual and resonance states are found in the second Riemann sheet. 
Here we call state in the second Riemann sheet above the threshold as resonance 
and one below the threshold as virtual state. 
With $s$-wave interaction of the quark and antiquark for the pseudoscalar mesons,
virtual states can be generated with a decay width.

Let us consider first the chiral limit with $m_{q} = m_{s} =0$, where 
it is easier to identify the dynamical symmetry breaking
by nonzero quark condensates.
There we have also the flavor SU(3) symmetry as $\vev{\bar qq} = \vev{\bar ss}$ 
and $M_{q}=M_{s}$. 
The effective potential per flavor, $\tilde V \equiv V(M)/N_{f}$, is 
given by a function of the dynamical quark mass $M$ for $M \ge 0$
\begin{equation}
   \tilde V(M)  = \tilde V(0) \sqrt{1 + \frac{M^{2}}{\Lambda^{2}}} - \frac34 M \vev{\bar qq} 
   - \frac{g_{S}}2 \vev{\bar qq}^{2} - \frac{g_{D}}3 \vev{\bar qq}^{3}, \label{eq:VeffCL}
\end{equation}
with $\tilde V(0) = - {3\Lambda^{4}}/({4\pi^{2}})$. 
From Eq.~(\ref{eq:SF}), one can easily see 
that $\vev{\bar qq}$ decreases monotonically  
as $M$ increases and approaches a constant gradually at large $M \gg \Lambda$
in the current regularization. With $\vev{\bar qq} = 0$ at $M=0$, one finds that
$\vev{\bar qq}$ has a negative value for $M>0$.
The vacuum condition is obtained by $\ptdel{\tilde V}{M}=0$ at $M=M_{q}$ as
\begin{equation}
   (-M_{q}-g_{S} \vev{\bar qq} - g_{D} \vev{\bar qq}^{2}) \left. \pdel{\vev{\bar qq}}{M} \right|_{M=M_{q}}=0.
   \label{eq:vc}
\end{equation}
Because one finds that  $\ptdel{\vev{\bar qq}}{M}$ has a nonzero negative value
while Eq.~\eqref{eq:vc} has a trivial solution with $M_{q}=0$,
nonrivial solutions of Eq.~\eqref{eq:vc} are obtained by solving the gap equation 
$ M_{q} + g_{S} \vev{\bar qq} + g_{D} \vev{\bar qq}^{2} = 0$.

We calculate the curvature of the effective potential $\tilde V(M)$ at the origin:
\begin{equation}
  \left. \pdel{^{2}\tilde V(M)}{M^{2}} \right|_{M=0} 
  = \frac{3\Lambda^{2}}{2\pi^{2}}(1 -  \frac{3\Lambda^{2}}{2\pi^{2}}g_{S}).
\end{equation}
This implies that for $g_{S} > g_{S}^\textrm{crit}$,
where the critical value is defined by 
$g_{S}^\textrm{crit} = 2\pi^{2}/(3\Lambda^{2})$, 
the effective potential has a local maximum at $M=0$ and 
the minimum value at $M=M_{q}$ with a finite $M_{q}$.
This is the standard scenario of the dynamical breaking of 
chiral symmetry in the NJL model.

Even though the effective potential has a local minimum at the origin 
with $g_{S} < g_{S}^\textrm{crit}$, if Eq.~\eqref{eq:vc} has a solution
with a finite $M_{q}$ and the effective potential has the minimum value
there, chiral symmetry is dynamically broken. 
That is possible with negative values of $g_{D}$ provinding 
\begin{align}
   \left. \pdel{^{2}\tilde V(M)}{M^{2}} \right|_{M=M_{q}} > 0, \qquad
   \tilde V(M_{q})& < \tilde V(0),
\end{align}
with a finite $M_{q}$ satisfying Eq.~\eqref{eq:vc}.
We call this situation anomaly driven symmetry breaking.

\begin{figure}
\centering
   \includegraphics[width=0.8\linewidth]{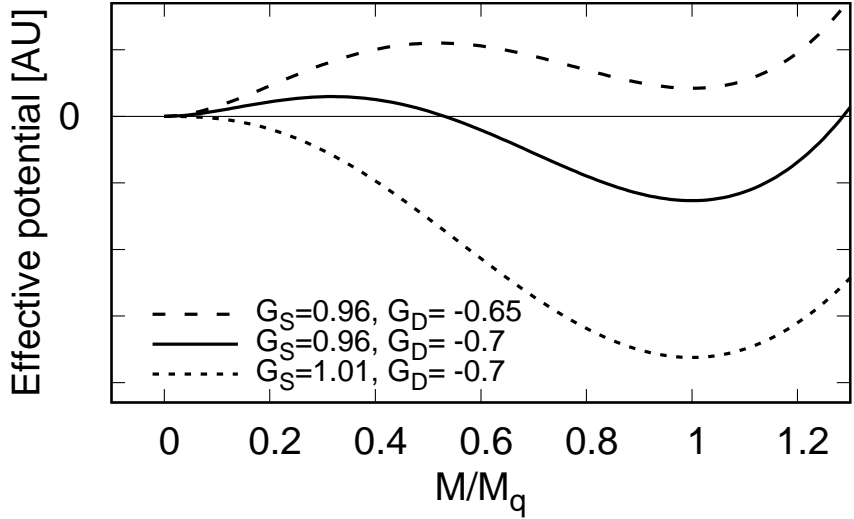}
\caption{Schematic plot of the effective potential calculated by the NJL model
in the chiral limit and the mean field approximation as a function of the dynamical quark mass $M$ 
scaled to the nontrivial solution $M_{q}$ of the gap equation. We have removed trivial constant terms for comparison. 
The calculation is done with $G_{S}=0.96,\ G_{D}=-0.65$ (dashed line), $G_{S}= 0.96,\ G_{D}=-0.7$ (solid line) 
and $G_{S}=1.01,\ G_{D}=-0.7$ (dotted line).  
See text for the details. } \label{fig:Veff}
\end{figure}

In Fig.~\ref{fig:Veff}, we show three examples of the effective potential 
as functions of the dynamical quark mass $M$ scaled to the nontrivial solution 
of the gap equation,~$M_{q}$. In the plot, trivial constant terms are removed
for comparison. 
Here we have introduced dimensionless coupling constants defined by
\begin{equation} \label{DLparam}
   G_{S}  = \frac{g_{S}}{g_{S}^{\textrm{crit}}} =  \frac{3\Lambda^{2}}{2\pi^{2}} g_{S}, \qquad
   G_{D}  = \frac{\Lambda g_{D}}{(g_{S}^{\textrm{crit}})^{2}} = \Lambda \left(\frac{3\Lambda^{2}}{2\pi^{2}}\right)^{2}g_{D}.
\end{equation}
The calculation is done with three examples;  
dashed line $(G_{S},G_{D})=(0.96,-0.65)$, solid line $(G_{S},G_{D})= (0.96,-0.7)$
and dotted line $(G_{S},G_{D})=(1.01,-0.7)$. 
For $G_{S}=1.01$, $g_{S}$ is larger than the critical value $g_{S}^\textrm{crit}$ 
and the effective action does not have a local minimum at  $M=0$, 
and thus chiral symmetry is broken dynamically without the anomaly term. 
In contrast, 
for $G_{S}<1$, the effective potential has a local minimum at $M=0$
and with sufficiently large $G_{D}$ in magnitude the effective potential
has the minimum at a nonzero quark mass. 
The latter situation is the anomaly driven breaking.

\begin{figure}
\centering
\includegraphics[width=0.8\linewidth]{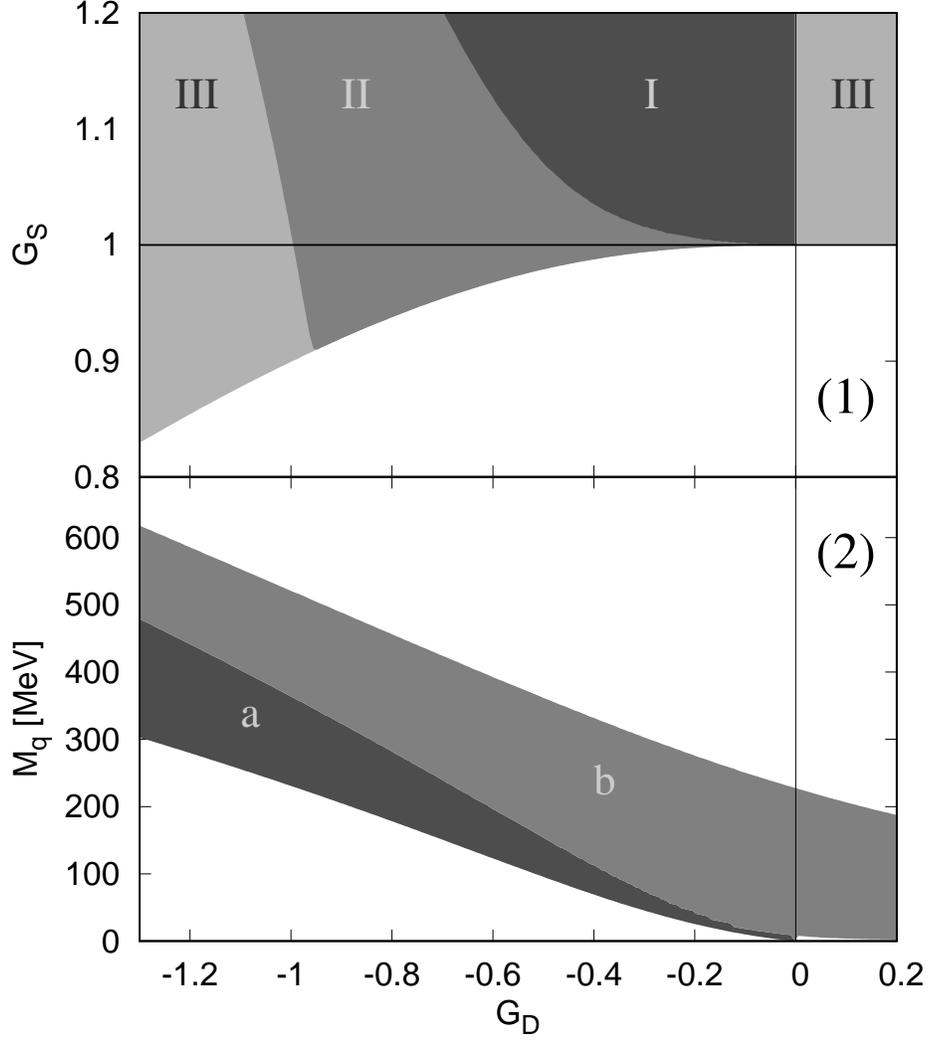}
\caption{
Upper (1):
Phase diagram for the dimensionless coupling constants $G_{D}$ and $G_{S}$ in the chiral limit. 
The dimensionless parameters are given in Eqs.~\eqref{DLparam}. 
The black and gray areas indicate pairs of the coupling constants 
where chiral symmetry is dynamically broken. 
In region I, the $\eta_{0}$ meson is found as a bound state, 
while it is a virtual state or a resonance in region II.
In region III, there is no solution for $\eta_{0}$. 
Lower (2):
The dynamical quark mass $M_{q}$ obtained with a parameter set $(G_{S},G_{D})$ 
in the chiral limit. The black area labeled by `a' denotes the dynamical 
quark mass obtained with the anomaly driven breaking $G_{S} < 1$, 
while the gray area labeled by `b' stands for $M_{q}$ obtained with 
the normal symmetry breaking $G_{S} \ge 1$. 
}  \label{fig:PhDMqCL}
\end{figure}

In order to see the condition of the dynamical breaking in more details, 
we show in Fig.~\ref{fig:PhDMqCL} (1) a phase diagram for the dimensionless model parameters
$(G_{D}, G_{S})$ with which the dynamical breaking of chiral symmetry takes place
with a finite quark mass $M_{q}$. 
The figure shows the dynamical breaking phase in the black and gray areas. 
As one sees, for $G_{D} <  0$,
chiral symmetry is dynamically broken,
even though the coupling constant $G_s$
is smaller than its critical value.

In Fig.~\ref{fig:PhDMqCL} (2), we show the dynamical quark masses obtained 
by a parameter set $(G_{S}, G_{D})$. It is shown that 
the stronger $G_{D}$ in magnitude provides the larger dynamical mass. 
In the black area labeled by `a', 
the dynamical quark mass is obtained with the anomaly driven breaking,
that is $G_{S}<1$, while in the gray area labeled by `b', it is obtained 
with the standard symmetry breaking $G_{S}\ge 1$. The figure shows
that the dynamical quark mass in the anomaly driven breaking is 
smaller than that in the standard symmetry breaking for each $G_{D}$. 

In Fig.~\ref{fig:PhDMqCL} (1), 
the properties of the $\eta_{0}$ meson are categorized by the gray scale. 
The black region labeled by~`\textrm{I}' represents where 
the $\eta_{0}$ meson is found as a bound state of a quark and an antiquark. 
In region \textrm{I}, only the usual chiral symmetry breaking 
is the case.
This is consistent with the linear $\sigma$ model where 
for the lighter $\eta_{0}$ masses the usual symmetry breaking 
takes place. 
The dark gray area specified by `\textrm{II}' in Fig.~\ref{fig:PhDMqCL}~(1) 
shows the region that the $\eta_{0}$ meson 
is found as a solution of Eq.~\eqref{eq:mass} 
in the second Riemann sheet, 
which is either a resonance appearing above the $2M_{q}$ threshold
or a virtual state below the threshold.
In region~\textrm{II}, the anomaly driven breaking can be the case. 
This implies that 
for a heavier $\eta_{0}$ mass the anomaly driven breaking can take place.
This is also consistent with the linear $\sigma$ model. 
In the gray areas labeled by `\textrm{III}', no solution of Eq.~\eqref{eq:mass}
for the $\eta_{0}$ meson is found in the region of
$\textrm{Re}\, m_{\eta_{0}} < 2\sqrt{M_{q}^{2} + \Lambda^{2}}$~\footnote{The loop functions $J(M)$ have a singularity at $\sqrt s = 2\sqrt{M^{2}+\Lambda^{2}}$.}
and $- \textrm{Im}\, m_{\eta_{0}} < 2M_{q}$.  
In particular, for $G_{D}=0$  
the $\eta_{0}$ meson would be found massless as a Nambu-Goldstone boson 
associated with the dynamical breaking of the U(3)$_{L} \otimes$U(3)$_{R}$
chiral symmetry, 
and for $G_D > 0$, 
the $\eta_{0}$ meson receives further 
attraction which makes $\eta_{0}$ unstable. Thus, we do not consider 
$G_D > 0$ for further calculations.

For the $\sigma_{0}$ mass, it is known that the $\sigma_{0}$ mass is equal to 
$2M_{q}$ in the chiral limit with $G_{D}=0$ in the ladder approximation~\cite{Nambu:1961tp,Hatsuda:1994pi}.
For $G_{D} < 0$ the anomaly term works attractively to the $\sigma_{0}$ channel. Thus,
the $\sigma_{0}$ meson is a bound state of the quark and antiquark for $G_{D} < 0$, while
the anomaly term with $G_{D}>0$ contribute to the $\sigma_{0}$ channel repulsively 
and the $\sigma_{0}$ is found to be a resonance state.

\begin{figure}
\centering
\includegraphics[width=0.8\linewidth]{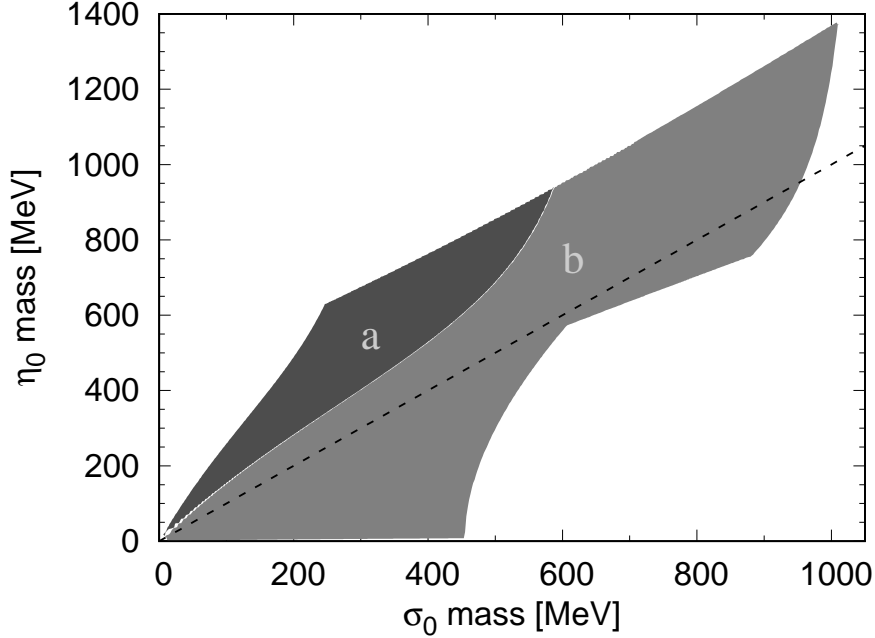}
\caption{
Correlation between the real parts of
the $\sigma_{0}$ and $\eta_{0}$ masses
obtained in the chiral limit with the parameter set $(G_{S}, G_{D})$ in region I and II of Fig.~\ref{fig:PhDMqCL}(1).
The dashed line stands for $m_{\eta_{0}}=m_{\sigma_{0}}$. 
The black area labeled by `a' denotes the meson masses 
obtained with the anomaly driven breaking $G_{S} < 1$, 
while the gray area labeled by `b' stands for $m_{\sigma_{0}}$ and $m_{\eta_{0}}$ 
obtained with the normal symmetry breaking $G_{S} \ge 1$.
}  
\label{fig:msig0meta0CL}
\end{figure}

In Fig.~\ref{fig:msig0meta0CL} we show the correlation of the real parts 
of the $\sigma_{0}$ and $\eta_{0}$ masses obtained by the parameter set $(G_{S}, G_{D})$
shown in region I and II of Fig.~\ref{fig:PhDMqCL}(1). The black area shows again the 
masses obtained in the anomaly driven breaking with $G_{S}<1$. This figure shows that
in the anomaly driven breaking the $\eta_{0}$ mass is always larger than that of 
the $\sigma_{0}$ meson. 
Furthermore, only the standard breaking can take place for large $\sigma_0$ masses. 
These are consistent with the linear $\sigma$ model. 
Because this is a chiral limit calculation, it is no problem even though 
the observed values of the $\eta^{\prime}$ and $f_{0}(500)$ masses are out of the area.

It is natural that due to the lack of the confinement 
mechanism in the NJL model the mesons generated dynamically in the scattering of a
quark and an antiquark have a decay width to fall apart quarks. 
Owing to the $s$-wave nature of the interaction between the quark and the antiquark for 
the pseudoscalar channel, the coupling of the pseudoscalar meson to the quarks does not 
have to vanish at the threshold and the decay width can be quite large for resonance states,
such as $\eta_{0}$. In contrast, for the scalar channel,
the coupling of the scalar meson to the quark and the antiquark should vanish at the threshold
and its decay width is suppressed around the threshold. Nevertheless, 
these widths are model artifacts coming from the lack of the confinement mechanism.

In the chiral limit, 
one has the Glashow-Weinberg relation~\cite{Glashow:1967rx} 
\begin{equation}
FG = - 2 \langle \bar qq \rangle,
\end{equation}
where the quantities $F$ and $G$ are defined by the matrix elements for the Nambu-Goldstone boson as
\begin{align}
\langle 0 | A_{\mu}^{a}(x) | \pi^{b} (p)\rangle &= i \delta^{ab} p_{\mu} F e^{-ip\cdot x}, \label{eq:decayconst} \\
\langle 0 | P^{a}(x) | \pi^{b}(p) \rangle &= \delta^{ab} G e^{-ip\cdot x},
\end{align}
with
the axial vector current $A_{\mu}^{a} = \bar q \gamma_{\mu} \gamma_{5} \frac{\lambda^{a}}{2} q$
and the pseudoscalar field $P^{a} = \bar q i\gamma_{5} \lambda^{a} q$. 
This is an exact and model-independent relation obtained as one of the low-energy theorems of chiral symmetry. 
The Glashow-Weinberg relation implies that 
the meson decay constant $F$ behaves like the quark condensate,
or equivalently the dynamical quark mass $M$ as seen in Eq.~\eqref{eq:fpiCL} of Appendix~\ref{sec:fpi}, where we  explain the calculation of the
meson decay constant in the NJL model.
We show also in Fig.~\ref{fig:FpiCL} 
the behavior of the meson decay constant obtained in the chiral limit with 
the parameter set $(G_{S},G_{D})$ same as in Fig.~\ref{fig:PhDMqCL}.

\begin{figure}
\centering
\includegraphics[width=0.8\linewidth]{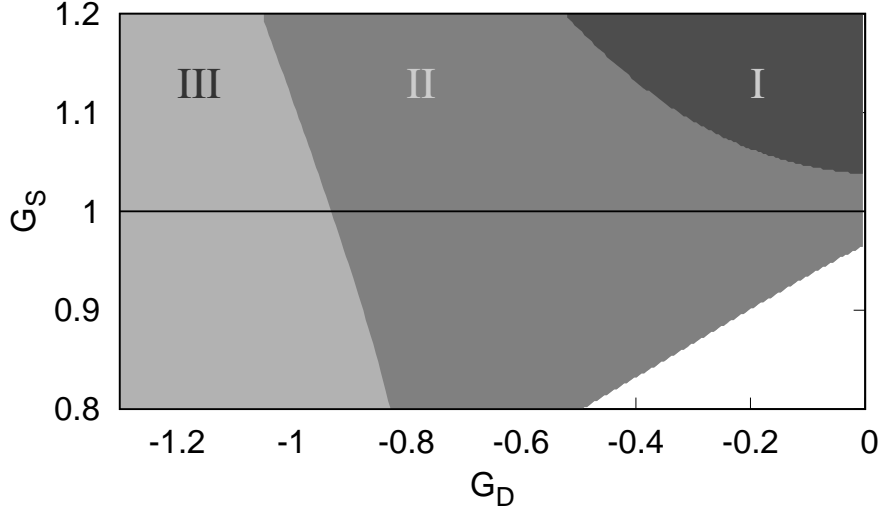}
\caption{
Same as in Fig.~\ref{fig:PhDMqCL}(1) calculated with the finite current quark masses. 
With the parameter set $(G_{S},G_{D})$ shown in the white area, the pion and kaon 
are not found with their physically observed masses as bound states of the dynamical quarks. 
} \label{fig:PhD}
\end{figure}

Let us now introduce the explicit chiral symmetry breaking 
with finite current quark masses. 
We determine the values of the current quark masses so as to reproduce the
isospin averaged masses of the pion and kaon as $m_{\pi} = 138.04$~MeV
and $m_{K}=495.65$~MeV, respectively, for each model parameter set 
$(G_{S},G_{D})$.
In order to survey a wide parameter region, 
we do not fix the values of the meson decay constants. 
We take the model parameters in the region of
\begin{equation}
   0.8 \le G_{S} \le 1.2, \qquad -1.3 \le G_{D} \le  0.0 .
   \label{eq:param}
\end{equation}
Within the above parameter region, we show the results when 
the pion and kaon are found to be bound state solutions
of Eq.~\eqref{eq:mass} with the observed masses. 
These parameter sets are shown in Fig.~\ref{fig:PhD} as the gray and black ares.
The values of the current quark masses are found to be 
distributed in the following ranges:
\begin{subequations}
\begin{alignat}{4}
   4.3~\textrm{MeV} < &\ m_{q}&\,  <& \ 5.8~\textrm{MeV},& \quad & \textrm{ave. }& 5.5~ \textrm{MeV},& \\
   120~\textrm{MeV} < &\ m_{s}& <& \ 170~\textrm{MeV},&  & \textrm{ave. }& 141~\textrm{MeV}.&
\end{alignat}
\end{subequations}
As seen in Fig.~\ref{fig:PhD}, while the anomaly driven symmetry breaking takes place
with wider parameter area than in the chiral limit, the $\eta_{0}$ meson 
becomes less stable. With larger $G_{D}$ in magnitude, the $\eta_{0}$ meson becomes
unbound.  
The white area in Fig.~\ref{fig:PhD} means that the pion and kaon cannot be 
reproduced with the physical masses as bound states of the dynamical quarks.

\begin{figure}
\centering
\includegraphics[width=0.8\linewidth]{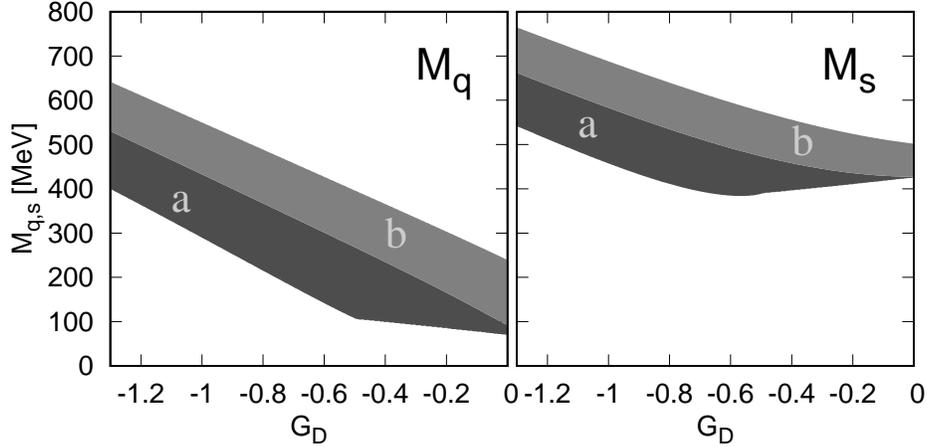}
\caption{
Dynamical quark masses, $M_{q}$ (left) and $M_{s}$ (right), calculated 
with a parameter set $(G_{S}, G_{D})$ and finite current quark masses. 
The black area labeled 
by `a' denotes the dynamical quark masses obtained with 
the anomaly driven breaking $G_{S} < 1$, 
while the gray area labeled by `b' stands for $M_{q}$ and $M_{s}$ obtained with 
the normal symmetry breaking $G_{S} \ge 1$. 
The current quark masses 
are determined so as to reproduce the pion and kaon masses. 
} \label{fig:mqms}
\end{figure}

We calculate the dynamical quark masses $M_{q}$ and $M_{s}$ for 
each parameter set $(G_{S}, G_{D})$
by the gap equations~\eqref{gapeq}. 
The obtained dynamical masses are shown in Fig.~\ref{fig:mqms}.
Again with the larger $G_{D}$ we obtain the heavier dynamical 
masses, and the dynamical masses are smaller in the anomaly 
driven breaking than in the standard breaking for each $G_{D}$. 
The behavior of the pion and kaon decay constants in the parameter region~\eqref{eq:param} 
is shown in Fig.~\ref{fig:fpifk} of Appendix~\ref{sec:fpi}.

\begin{figure}
\centering
   \includegraphics[width=0.8\linewidth]{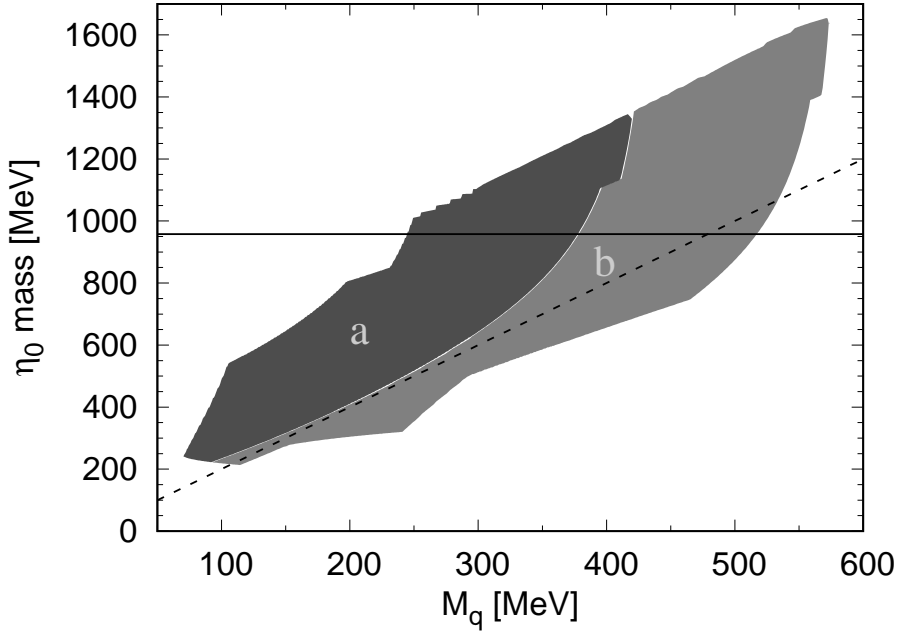}
\caption{Correlation between the values of the real part of
the $\eta_{0}$ mass and the dynamical quark mass $M_{q}$. The area indicated by `a' 
is where the values of the $\eta_{0}$ masses 
are obtained with $G_{S}<1$,
in which the anomaly driven breaking takes place.
The area indicated by `b' is where the values of
the $\eta_{0}$ masses are obtained 
with $G_{S}>1$,
in which the standard symmetry breaking takes place.
The horizontal solid line shows the observed $\eta^{\prime}$ mass, while the dashed 
line expresses $2 M_{q}$. The $\eta_{0}$ mesons found above the dashed line are resonance states.  
} \label{fig:metap}
\end{figure}

With the dynamical quark masses obtained above,
we calculate the $\eta_{0}$ meson mass
by solving Eq.~\eqref{eq:mass}
with the parameters in the range of Eq.~\eqref{eq:param}.
The $\eta_{0}$ mass is looked for in the region of 
$\textrm{Re}\, m_{\eta_{0}} < 2 \sqrt{M_{q}^{2} + \Lambda^{2}}$ and $-\textrm{Im}\, m_{\eta_{0}} < 2M_{s}$.
In Fig.~\ref{fig:metap}, we show the correlation between 
the real part of the $\eta_{0}$ mass 
and the dynamical quark mass $M_{q}$. First of all, we find that 
the $\eta_{0}$ mass is strongly dependent on the model parameters.
We specify the $\eta_{0}$ mass obtained with $G_{S}<1$,
where the anomaly driven breaking takes place, as the black area 
labeled by `a', while the $\eta_{0}$ mass obtained with
$G_{S}>1$, where the standard symmetry breaking takes place,
is shown as the gray area labeled by `b'. 
Figure~\ref{fig:metap} shows that in the anomaly driven breaking 
the $\eta_{0}$ meson is found as a resonance
with a relatively smaller dynamical quark mass.
In the standard breaking, the $\eta_{0}$ meson can be found as a bound state. 
It is also interesting to note that the $\eta_{0}$ mass looks proportional to the dynamical 
quark mass. This implies that the $\eta_{0}$ meson is a bound (or resonance) state of the dynamical 
quark and antiquark like the $\sigma_{0}$ meson not a Nambu-Goldstone boson. Thus, the constituents 
of the $\eta_{0}$ meson are the dynamical quarks and the $\eta_{0}$ mass scales to the dynamical quark mass.

\begin{figure}
\centering
   \includegraphics[width=0.8\linewidth]{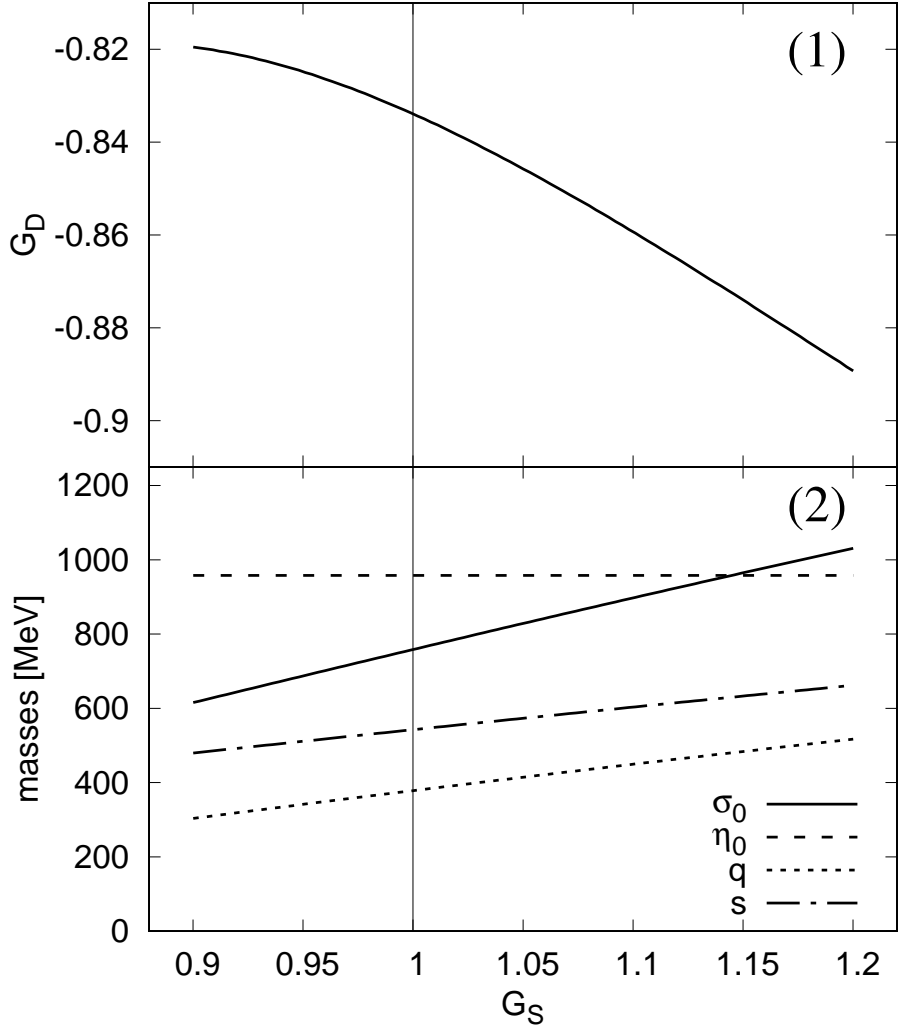}
\caption{
Upper (1): Determined $G_{D}$ parameter so as to reproduce the $\eta^{\prime}$ mass
for each $G_{S}$. Lower (2): Masses of the $\sigma_{0}$ meson, the $\eta_{0}$ meson (input) 
and dynamical quark masses calculated with the determined $G_{D}$ as functions of $G_{S}$. }
\label{fig:sigma}
\end{figure}

Now let us fix the value of the $G_{D}$ parameter so as to reproduce
the $\eta^{\prime}$ mass. For each $G_{S}$ we determine the $G_{D}$ parameter
so that the real part of the $\eta_{0}$ mass is obtained to be 
the observed $\eta^{\prime}$ mass $m_{\eta^{\prime}}=957.78$ MeV. 
The determined $G_{D}$ is shown in Fig.~\ref{fig:sigma}~(1) as a function of $G_{S}$.
In Fig.~\ref{fig:sigma}~(2), we show the corresponding $\sigma_{0}$ mass together with
the dynamical quark masses. The $\sigma_{0}$ mass is found to be very closed 
to $2M_{q}$. The figure shows that 
\begin{subequations}
\label{eq:sigmamass}
\begin{align}
   m_{\sigma_{0}}& < 760\ \textrm{ MeV}, \qquad  \textrm{ for}\quad G_{S} < 1, \\
   m_{\sigma_{0}}& > 760\ \textrm{ MeV}, \qquad  \textrm{ for}\quad G_{S} > 1.   
\end{align}
\end{subequations}
This means that 
if $m_{\sigma_{0}} < 760$ MeV the anomaly driven 
breaking is the case. 
This is consistent with the linear $\sigma$ model.

It is interesting to mention that the pion and kaon decay constants calculated with the fixed values of $G_{D}$
in the region $0.9 < G_{S} < 1.2$ run from 87~MeV to 98~MeV and 94~MeV to 98~MeV, respectively, 
which are consistent with the observed values even though the kaon decay constant is a bit underestimated. 
These values are calculated directly from the matrix element 
of the axial vector current. The meson decay constant can be also evaluated through the Gell-Mann Oakes Renner
relation, which is valid in the first order of the quark mass. We confirm that the decay constants obtained by 
the Gell-Mann Oakes Renner relation agree with the direct calculation within 1\%. 
We also perform a calculation in which the cut-off parameter $\Lambda$ is allowed to be changed 
and is fitted so as to reproduce the observed pion decay constants. 
The result is shown in Fig.~\ref{fig:sigma_fpi_fit} and is found to be essentially same 
as in Fig.~\ref{fig:sigma}. Equations~\eqref{eq:sigmamass} are also obtained
in the case of Fig.~\ref{fig:sigma_fpi_fit}.
Because the pion decay constant calculated with the fixed cut-off already reproduces 
consistently the observed value, the determined cut-off is almost constant against $G_{S}$. 
The detailed discussions are given in Appendix~\ref{sec:fpi}. 

\section{Summary}

We have shown that with sufficiently large anomaly contributions
chiral symmetry is broken spontaneously even if chiral symmetry 
is not broken without the anomaly term, 
both in the linear $\sigma$ model and the NJL model. 
This is not a surprising result, because the anomaly term can play a role for 
the dynamical breaking of chiral symmetry.

In the linear $\sigma$ model,
such an anomaly driven symmetry breaking is the case
for the $\sigma_{0}$ mass within
$ \frac19 m_{\eta_{0}}^{2} < m_{\sigma_{0}}^{2} <  \frac13 m_{\eta_{0}}^{2}$ 
compared with the $\eta_{0}$ mass
in the chiral limit.  This implies that
the $\sigma_{0}$ meson should be light. In other words,
the $\eta_{0}$ mass should be larger than the mass scale of 
the chiral symmetry breaking, which may be the $\sigma_{0}$ mass. 
For the standard chiral symmetry breaking, 
we have found $m_{\eta_{0}}^{2} < 3 m_{\sigma_{0}}^{2}$ in the chiral limit,
which suggests that the $\sigma_{0}$ meson should be heavy.
With the explicit breaking 
included, the condition for the anomaly driven symmetry breaking
becomes more quantitative, that
is $m_{\sigma_{0}} < 840$ MeV in the linear $\sigma$
model
reproducing the $\pi$, $\eta_{0}$ and $\eta_{8}$ masses 
with the ratio $f_K/f_\pi$.
If the anomaly driven breaking
should not take place, this would give us the lower 
limit of the $\sigma_{0}$ mass. 

In the NJL model also, the anomaly driven breaking takes place
with sufficiently large anomaly contributions. It is interesting that,
in the chiral limit, where the phase boundary for the dynamical 
symmetry breaking is clearly defined, the anomaly driven breaking 
is not the case when the $\eta_{0}$ meson is found as a 
bound state of the quark and antiquark. 
This means that 
for the anomaly driven breaking the $\eta_{0}$ meson 
should have a larger mass than the typical energy scale of 
the chiral symmetry breaking.
This is consistent with 
the finding in the linear sigma model. 
With the explicit breaking, we have found also that 
the observed $\eta^{\prime}$ mass is reproduced 
with relatively small masses of the dynamical quark in the anomaly driven breaking, 
while in the standard symmetry breaking,
to reproduce the $\eta_{0}$ mass, larger dynamical masses are necessary. 
With the anomaly term fixed by the observed $\eta^{\prime}$ mass,
we have found that the anomaly driven symmetry breaking 
is the case with $m_{\sigma_{0}} < 760$ MeV.
This is also consistent with 
the linear sigma model. 
We note that, in both models at the chiral limit, the chiral symmetric vacuum with 
$\vev{\sigma_{0}}=0$, or $\vev{\bar qq} = 0$,
provides the local minimum. In other words, it corresponds to a meta-stable state, which indicates strong first order phase transition at non-zero temperature and/or density.

Both chiral effective theories conclude that, in the standard symmetry breaking,
the $\sigma_{0}$ meson regarded as the quantum fluctuation of 
the chiral condensate have a relatively large mass than a order of 800 MeV,
while the $\sigma_{0}$ mass
can be smaller in the anomaly driven symmetry breaking.
Since the anomaly-driven breaking of the chiral symmetry  provides a light scalar meson $\sigma_0$ which is a singlet under SU(3)$_{\rm flavor}$ symmetry, the nonet scalar mesons including the $\sigma_0$ might be able to explain the spectra of light scalar mesons below 1\,GeV. The analysis of this line with including the effect of mixing between $\sigma_0$ and an isospin zero member of the octet is done in Ref.~\cite{KHMJ}. 

\section*{Acknowledgements}

The work of D.J.\ was partly supported by Grants-in-Aid for Scientific Research from JSPS (17K05449, 21K03530).
The work of M.H. is supported in part by 
JPSP KAKENHI Grant Numbers 16K05345 and 20K03927.

\appendix
\section{Calculation details of the NJL model}

In this section we show the detailed calculations of the formulae using in the NJL model.
The loop integrals are performed with the three momentum cutoff $\Lambda$.

The right hand side of Eq.~\eqref{qq} is calculated as
\begin{align}
\lefteqn{
   \left\langle \bar{q} q \right\rangle = - i N_c {\lim_{x \to 0}} \textrm{ Tr} [S_{F}(x)]   
   = - i N_c \int \frac{d^{4}q}{(2\pi)^{4}} \textrm{ Tr}\left[ \frac{1}{\diracslash{p} - M } \right] }\nonumber \\&
   = -  \frac{N_c M}{\pi^{2}} \int_{0}^{\Lambda} \frac{p^{2}}{\sqrt{p^{2} + M^{2}}}dp
    = - \frac{N_c M\Lambda^{2}}{2\pi^{2}} 
  \left(\sqrt{1+\frac{M^{2}}{\Lambda^{2}}} 
  - \frac{M^{2}}{\Lambda^{2}} \log \frac{\Lambda + \sqrt{ \Lambda^{2} + M^{2}}}{M} \right).
  \label{eq:SF}
\end{align}

The first term of the effective potential \eqref{eq:Veff} is obtained as
\begin{align}
  \lefteqn{
  i\int \frac{d^{4}p}{(2\pi)^{4}}
   \textrm{Tr} \left[ \log(\diracslash{p}-M)+ \frac{\diracslash{p}-m}{\diracslash{p} - M} \right] 
   } \nonumber \\
   =&  - \frac{\Lambda^{4}}{8\pi^{2}} \left[\left(2+\frac{M^{2}}{\Lambda^{2}} \right) \sqrt{1 +\frac{M^{2}}{\Lambda^{2}}}
   - \frac{M^{4}}{\Lambda^{4}} \log \frac{\Lambda+ \sqrt{\Lambda^{2} +M^{2}}}{M}
  \right]  - \frac{M-m}{N_{c}} \vev{\bar qq},  
  \label{eq:log}
\end{align}
where we have removed terms independent of $M$. 

The loop functions for the pseudoscalar and scalar channels at rest
$P = (\sqrt s, 0,0,0)$ are given for a bound state $\sqrt s \le 2 M$
by
\begin{align}
  J_{\textrm{PS}}(\sqrt s; M) 
  &=N_{c} i \int \frac{d^{4}p}{(2\pi)^{4}} 
  \textrm{Tr}\left[i\gamma_{5}\frac{1}{\diracslash{p} - M} i\gamma_{5} \frac{1}{\diracslash{p} - \Pslash - M}\right]
  \nonumber \\
  &= \frac{N_{c}\Lambda^{2}}{4\pi^{2}} \left[2 \sqrt{1+\frac{M^{2}}{\Lambda^{2}}} 
  - \frac{2M^{2}-s}{\Lambda^{2}} \log \frac{\Lambda+\sqrt{\Lambda^{2}+M^{2}}}{M} 
 \right. \nonumber \\ & \quad \left.
  - \sqrt{\frac{s(4M^{2}-s)}{\Lambda^{4}}} \tan^{-1} \sqrt{\frac{s}{4M^{2} -s}} \sqrt{\frac{\Lambda^{2}}{\Lambda^{2}+M^{2}}}\right],
  \label{eq:JPS}
\end{align}
and
\begin{align}
  J_{\textrm{S}}(\sqrt s;M)
  &=N_{c} i \int \frac{d^{4}p}{(2\pi)^{4}} 
  \textrm{Tr}\left[\frac{1}{\diracslash{p} - M}\frac{1}{\diracslash{p} - \Pslash - M}\right]
  \nonumber \\
  &= \frac{N_{c}\Lambda^{2}}{4\pi^{2}} \left[2 \sqrt{1+\frac{M^{2}}{\Lambda^{2}}} 
  - \frac{6M^{2}-s}{\Lambda^{2}} \log \frac{\Lambda+\sqrt{\Lambda^{2}+M^{2}}}{M} 
 \right. \nonumber \\ & \quad \left.
  + \sqrt{\frac{(4M^{2}-s)^{3}}{s\Lambda^{4}}} \tan^{-1} \sqrt{\frac{s}{4M^{2} -s}} \sqrt{\frac{\Lambda^{2}}{\Lambda^{2}+M^{2}}}\right],
  \label{eq:JS}
\end{align}
respectively.
For a kaon, different quarks are in the loop:
\begin{align}
J_{\textrm{PS},K}(\sqrt s;M_{1},M_{2})
=& N_{c} i \int \frac{d^{4}p}{(2\pi)^{4}} 
  \textrm{Tr}\left[i\gamma_{5}\frac{1}{\diracslash{p} - M_{1}} i\gamma_{5} \frac{1}{\diracslash{p} - \Pslash - M_{2}}\right] \nonumber \\
=& \frac{N_{c} \Lambda^{2}}{4\pi^{2}} 
  \left\{ \sqrt{1+\frac{M_{1}^{2}}{\Lambda^{2}}} - \frac{M_{1}^{2} + M_{2}^{2} -s}{2\Lambda^{2}} 
  \log \frac{\Lambda+\sqrt{\Lambda^{2}+M_{1}^{2}}}{M_{1}}
  \right. \nonumber \\
  &- \frac{(M_{1}-M_{2})^{2}}{\Lambda^{2}} \frac{\omega_{1}}{\sqrt s} \log \frac{\Lambda+\sqrt{\Lambda^{2}+M_{1}^{2}}}{M_{1}}
  \nonumber \\ 
  & \left. - \frac{ s - (M_{1} - M_{2})^{2}}{\Lambda^{2}}
   \frac{\eta}{\sqrt s} \tan^{-1} \frac{\omega_{1}}{\eta} \sqrt{\frac{\Lambda^{2}}{\Lambda^{2}+M_{1}^{2}}} \right\}
  + (M_{1} \leftrightarrow M_{2}),
 \label{eq:JK}
\end{align}
where we have defined 
\begin{align}
 & \omega_{1} = \frac{s + M_{1}^{2} - M_{2}^{2}}{2\sqrt s}, \quad
  \omega_{2} = \frac{s + M_{2}^{2} - M_{1}^{2}}{2\sqrt s}, \\
 & \eta = \frac{\sqrt{[(M_{1}+M_{2})^{2} - s][s- (M_{1}-M_{2})^{2}]}}{2\sqrt s}.
\end{align}

The loop functions for each channel are
\begin{subequations}
\label{loopfun}
\begin{align}
  J_{\eta_{0}} & =  \frac13 ( 2J_{\textrm{PS}}(\sqrt s; M_{q}) + J_{\textrm{PS}}(\sqrt s; M_{s})),\\
  J_{\pi} &= J_{\textrm{PS}}(\sqrt s; M_{q}) ,\\
  J_{K} &= J_{\textrm{PS},K}(\sqrt s; M_{q},M_{s}) ,   \\
  J_{\eta_{8}} & =   \frac13 ( J_{\textrm{PS}}(\sqrt s; M_{q}) + 2J_{\textrm{PS}}(\sqrt s; M_{s})),\\
  J_{\sigma_{0}} & =  \frac13 ( 2J_{\textrm{S}}(\sqrt s; M_{q}) + J_{\textrm{S}}(\sqrt s; M_{s})).
\end{align}
\end{subequations}

The interaction kernels for the mesonic channels are 
\begin{subequations}
\label{kernel}
\begin{align}
  K_{\eta_{0}} &  = g_{S} - \frac23 g_{D} (2\vev{\bar qq} + \vev{\bar ss}),\\
  K_{\pi} &= g_{S} + g_{D} \vev{\bar ss} ,\\
  K_{K} &
  = g_{S} + g_{D} \vev{\bar qq}, 
  \\
  K_{\eta_{8}} & = g_{S} + \frac13 g_{D} (4 \vev{\bar qq} - \vev{\bar ss}), \\
  K_{\sigma_{0}}& = g_{S} + \frac23 g_{D} (2\vev{\bar qq} + \vev{\bar ss}).
\end{align}
\end{subequations}

\section{Discussion on the meson decay constants in the NJL model}
\label{sec:fpi}

The meson decay constant is defined by Eq.~\eqref{eq:decayconst} and 
is calculated in the NJL model as
\begin{equation}
   F = -i \sqrt{2R} J_{A}(m),
\end{equation}
where $m$ is the meson mass and $R$ is the residue of the $T$ matrix at $s=m^{2}$.
The loop function with the axial current, $J_{A}$, is defined by 
\begin{equation}
   P^{\mu}J_{A}(P)  \equiv i N_{c} \int \frac{d^{4}p}{(2\pi)^{4}} 
   \textrm{Tr}\left[\frac{\gamma^{\mu}\gamma_{5}}2 \frac{1}{\diracslash{p}-M} 
   i \gamma_{5} \frac{1}{\diracslash{p} - \Pslash - M}\right], 
\end{equation}
and is calculated at the rest frame $P=(\sqrt s, 0,0,0)$ as 
\begin{align}
   J_{A}(\sqrt{s}) &=- \frac{iN_{c}\Lambda}{4 \pi^{2}}\left[
    \frac M \Lambda \log\frac{\Lambda + \sqrt{\Lambda^{2}+M^{2}}}{M}
     \nonumber \right. \\ & \left. \qquad
     -\frac M \Lambda \sqrt{\frac{4M^{2}-s}{s}}
     \tan^{-1}\sqrt{\frac s{4M^{2}-s}} \sqrt{\frac{\Lambda^{2}}{\Lambda^{2}+M^{2}}}
   \right].
\end{align}
For kaon, the loop function with different quark masses reads
\begin{align}
   \lefteqn{J_{A,K}(\sqrt{s}) =} & \nonumber \\& \quad 
   -\frac{iN_{c}\Lambda}{4\pi^{2}}
  \left\{
  \frac{\Lambda(M_{2}-M_{1})}{2 s} \sqrt{1 + \frac{M_{1}^{2}}{\Lambda^{2}}} 
   +\frac{M_{1}}\Lambda \frac{\omega_{1}}{\sqrt s} 
      \log \frac{\Lambda + \sqrt{\Lambda^{2} + M_{1}^{2}}}{M_{1}}
    \right. \nonumber \\ & \left. \quad
   + \frac{M_{1}-M_{2}}\Lambda \frac{\eta^{2}-\omega_{1}^{2}}{2s}  
      \log \frac{\Lambda + \sqrt{\Lambda^{2} + M_{1}^{2}}}{M_{1}}
   \nonumber \right. \\ & \left. \quad
     + \frac{M_{1}+M_{2}}\Lambda \frac{(M_{1}-M_{2})^{2}-s}{s} \frac{\eta}{2\sqrt s} \tan^{-1} \frac{\omega_{1}}{\eta} \sqrt{\frac{\Lambda^{2}}{\Lambda^{2}+M_{1}^{2}}}\right\}
     +  (M_{1} \leftrightarrow M_{2}).
\end{align}
The residue $R$ is calculated as
\begin{equation}
  R \equiv \lim_{s \to m^{2}} (s-m^{2}) T = - \frac{1}{\left.\frac{dJ}{ds}\right|_{s=m^{2}}},
\end{equation}
where we have used l'Hospital theorem and the fact that the $T$-matrix is given
from Eq.~\eqref{BSeq} as $T=K/(1-KJ)$ with constant $K$. 
The meson decay constant in the chiral limit
is calculated as
\begin{equation}
   F = M \sqrt{\frac{N_{c}}{2\pi^{2}}  \left(\sqrt{\frac{\Lambda^{2}}{\Lambda^{2}+M^{2}}} 
   - \log \frac{\Lambda+\sqrt{\Lambda^{2}+M^{2}}}{M}\right)}. \label{eq:fpiCL}
\end{equation}

\begin{figure}
\centering
\includegraphics[width=0.6\linewidth]{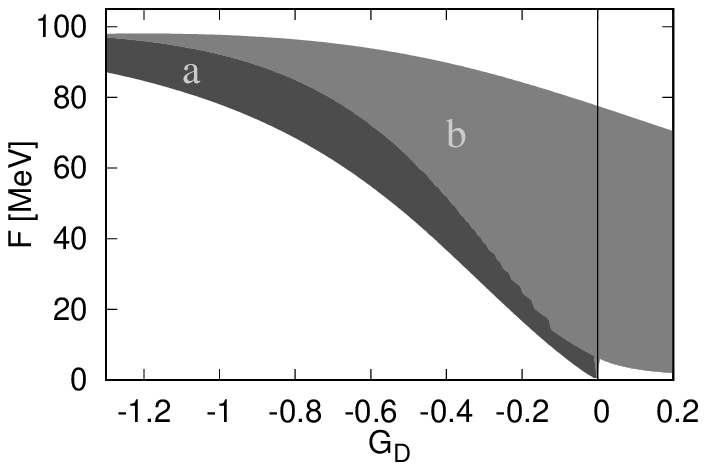}
\caption{
The meson decay constant $F$ obtained in the chiral limit 
for the parameter set $(G_{S},G_{D})$ same as in Fig.~\ref{fig:PhDMqCL}. 
The black area labeled by `a' denotes the meson decay constant 
obtained with the anomaly driven breaking $G_{S} < 1$, 
while the gray area labeled by `b' stands for $F$ obtained with 
the normal symmetry breaking $G_{S} \ge 1$. }
\label{fig:FpiCL}
\end{figure}

In Fig.~\ref{fig:FpiCL}, we show the meson decay constant obtained in the chiral limit 
with the parameter set $(G_{S}, G_{D})$ same as in Fig.~\ref{fig:PhDMqCL}.  
In the black area labeled by `a', 
the meson decay constant is obtained with the anomaly driven breaking, 
while in the gray area labeled by `b', it is obtained 
with the standard symmetry breaking. The figure shows
that the meson decay constant in the anomaly driven breaking is 
smaller than that in the standard symmetry breaking for each $G_{D}$, as seen 
in the dynamical quark mass. 

\begin{figure}
\centering
\includegraphics[width=0.8\linewidth]{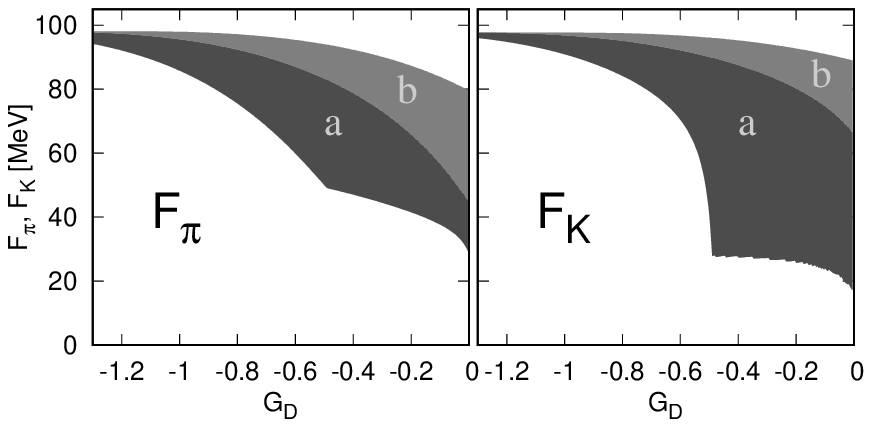}
\caption{
Meson decay constants, $F_{\pi}$ (left) and $F_{K}$ (right), calculated 
with the parameter set $(G_{S}, G_{D})$ same as in Fig.~\ref{fig:mqms} 
and finite current quark masses. 
The black area labeled 
by `a' denotes the meson decay constants obtained with 
the anomaly driven breaking $G_{S} < 1$, 
while the gray area labeled by `b' stands for $F_{\pi}$ and $F_{K}$ obtained with 
the normal symmetry breaking $G_{S} \ge 1$. 
The current quark masses 
are determined so as to reproduce the pion and kaon masses. 
} 
\label{fig:fpifk}
\end{figure}

{In Fig.~\ref{fig:fpifk}
we calculate the meson decay constants of pion and kaon, $F_{\pi}$ and $F_{K}$, 
with finite quark masses for the parameter set $(G_{S}, G_{D})$ given 
by Eq.~\eqref{eq:param}. 
Again we find that the meson decay constants are smaller in the anomaly 
driven breaking than in the standard breaking for each $G_{D}$. 
If one fixes the pion decay constant to the observed value, one finds correlation 
between the parameters $G_{S}$ and $G_{D}$. }

\begin{figure}
\centering
   \includegraphics[width=0.6\linewidth]{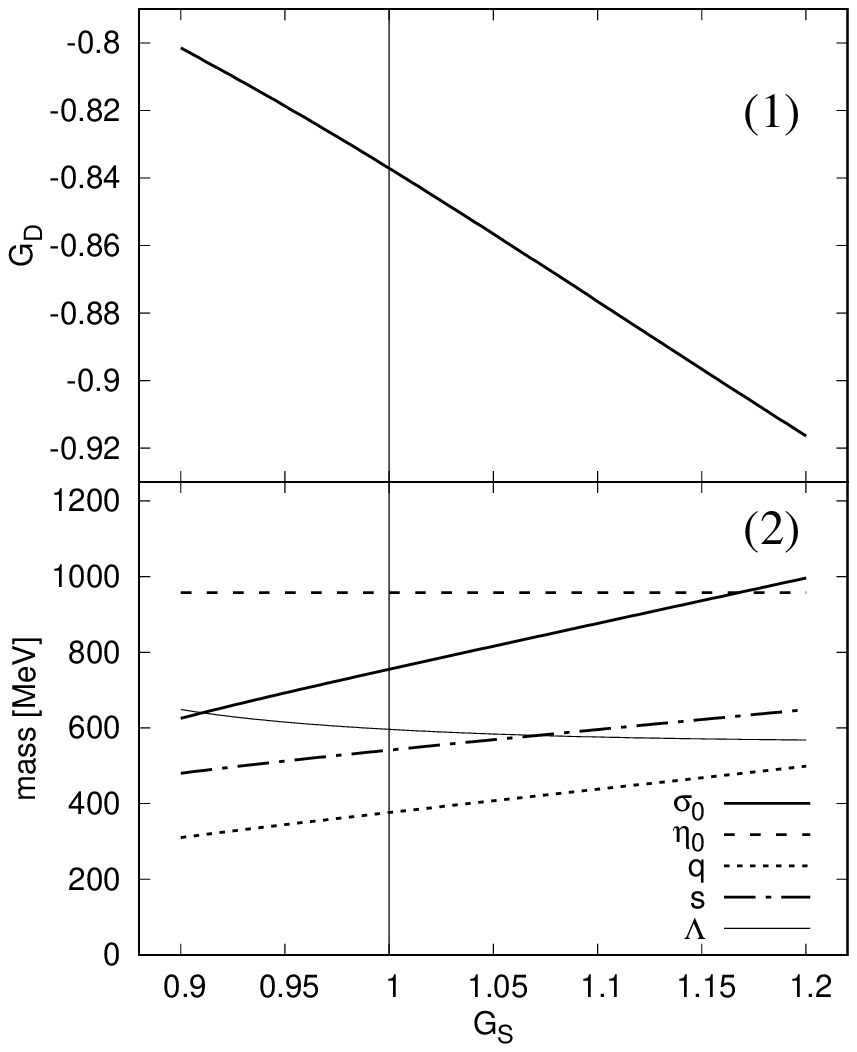}
\caption{
Same as in Fig.~\ref{fig:sigma} but calculation with
the cut-off parameter $\Lambda$ fitted so as to reproduce the pion decay constant
for each $G_{S}$.
Upper (1): Determined $G_{D}$ parameter so as to reproduce the $\eta^{\prime}$ mass
for each $G_{S}$. Lower (2): Masses of the $\sigma_{0}$ meson, the $\eta_{0}$ meson (input) 
and dynamical quark masses calculated with the determined $G_{D}$ as functions of $G_{S}$. 
The determined cut-off parameter $\Lambda$ is also shown. 
}
\label{fig:sigma_fpi_fit}
\end{figure}

In Fig.~\ref{fig:sigma_fpi_fit} we show the result in which the parameter $G_{D}$ 
and the cut-off $\Lambda$ are determined so as to obtain the $\eta_{0}$ mass 
to be the observed $\eta^{\prime}$ mass $m_{\eta^{\prime}}=957.78$ MeV
and the pion decay constant to be $F_{\pi} = 92.2$ MeV for each $G_{S}$. 
The determined $G_{D}$ is shown in Fig.~\ref{fig:sigma_fpi_fit} (1) and found 
to be almost same as the calculation with the fixed cut-off given in Fig.~\ref{fig:sigma} (1). 
Figure~\ref{fig:sigma_fpi_fit} (2) shows the $\sigma_{0}$ masses,
the dynamical quark masses, $M_{q}$ and $M_{s}$,
and the determined cut-off $\Lambda$ for each $G_{S}$. The masses are again 
consistent with Fig.~\ref{fig:sigma} (2). The determined cut-off is almost constant against 
$G_{S}$. 

\end{document}